\documentstyle[aps,12pt,epsf]{revtex}
\begin{document}
\title{Coherent and chaotic properties of nuclear pairing}
\author{Alexander Volya, Vladimir Zelevinsky, and B. Alex Brown}
\address{Department of Physics and Astronomy and \\
National Superconducting Cyclotron Laboratory, \\
Michigan State University,
East Lansing, Michigan 48824-1321, USA}
\maketitle
\begin{abstract}
\baselineskip 14pt
The properties of the pairing interaction in the shell model framework are
considered with the aid of the exact numerical solution utilizing the quasispin
symmetry. We emphasize the features which are out of reach for the usual
approximate techniques based on the BCS approach supplemented by the random
phase approximation treatment of pair vibrations, especially in the region of
weak pairing where the BCS+RPA theory fails. For the first time chaotic aspects
of the mixing generated by the pairing interaction are studied. The level
repulsion and large information entropy of the eigenstates coexist with the 
absence of thermalization of single-particle motion. The full spectrum of pair
vibration in average displays the spin dependence similar to that for a 
rigid rotor.
\end{abstract}

\label{pure}
\section{Introduction}

Many properties of atomic nuclei are essentially 
governed by pairing interactions. Here and below 
we call ``pairing" the part of residual
interaction which scatters the particles between the pairs of time-conjugate
single-particle orbits. The isoscalar pairing, which may be important 
for $N\approx Z$ nuclei far from stability, will not be discussed here (at a
single spherical $j$-level such $T=0$ pairing would require pairs with $J=1$),
although it also can be studied in a similar manner.

Pairing is often viewed as a simple and the most regular part of the nuclear 
interaction. In a low-lying region of the spectrum 
it forms a pair condensate which strongly influences all nuclear properties
\cite{Bel,BM}. 
According to the standard BCS description borrowed from the macroscopic theory
of superconductivity,
the excitation of the system breaks pairs removing them from the interaction
domain and blocking the scattering phase space for remaining pairs. Another,
collective
mode of excitation implies the redistribution of the pairs with no breaking.
As a result, coherent pair 
vibrations of the condensate are possible.
At some excitation energy (or temperature) there occurs a sharp second order
phase transition to a
normal heated Fermi-liquid where the pairing effects are usually neglected,
again in the spirit of the BCS theory of macroscopic
superconductors.
However, closer consideration of the problem for finite systems such as nuclei
reveals that there exist  
nontrivial effects resulting from pairing that persist through the entire
shell-model spectrum. In nuclei the mesoscopic nature of the system smoothens 
all transitional phenomena resulting in more gradual change of properties.

Another (usually ignored) aspect of the pairing problem is that,
along with the very coherent nature of
the ground state, pairing brings in a certain degree of incoherent mixing. 
Our goal below is to study the pairing problem in full in order to present a
consistent picture of the
physical effects associated with the pairing interaction.
For a broad range of energies 
we study the thermodynamical properties, entropy, thermalization and
mixing of simple configurations as well as coherent pair excitations,
and average moments of inertia. A thorough
consideration of pairing should precede the detailed study of the interplay of
the pairing with other components of the residual interaction (the shell model
analysis reveals \cite{big} unexpected results of this interplay).

We study pairing in the framework of the realistic spherical nuclear shell 
model, which provides
a perfect testing ground for discussing the phenomena of interest. 
There exist a number of approximate methods  for curing
the well known shortcomings of the BCS approach for finite
systems. 
However they cannot be used for our purpose. Pairing fluctuations
extending far below the BCS phase transition cannot be 
treated with the usual approximate techniques.
The spectrum of excited states, 
including those at high energy, is also
out of reach for the most of approximations. 
Having this situation in mind, 
we consider the properties of the exact solution of a nuclear 
many-body problem with the pairing interaction \cite{EP} using an
algorithm based on the quasispin method \cite{kerman61}. Earlier we have shown
\cite{EP} 
that such an exact numerical solution is convenient, practically simple and can
be combined with the new approaches for taking into account other parts of the
residual interaction.

\section{The model}

We consider a general description of the finite Fermi-system in the 
framework of the spherical shell model. Our space includes
a set of $j$-levels (of capacity $\Omega_j=2j+1$) with single-particle 
energies $\epsilon_{j}$. The method \cite{kerman61,EP}
used here for solving the pairing problem exactly is based on the
quasispin algebras  present in the pairing Hamiltonian
\begin{equation}
H=\sum_{j}\epsilon_{j} N_{j} \,+\,
\sum_{j,\,j'} {\cal V}_{jj'}\,P^{\dagger}_j 
P_{j'}.                                                    \label{BCS:ham}
\end{equation}
Here $P^{\dagger}_j$ and $P_j$ are pair creation and annihilation operators of
particles on a $j$-level:
\begin{equation}
P_j=\frac{1}{2}\,\sum_{m}\, {\tilde a_{j\,m}}
a_{j\,m}\,,\quad
P^{\dagger}_j=\left (P_j \right)^\dagger=\frac{1}{2}\,\sum_{m}\,
a^{\dagger}_{j\,m} {\tilde a^{\dagger}_{j\,m}}\,,
\end{equation}
where the tilde refers to time-conjugate states,
$\tilde{a}_{j\,m}\equiv (-)^{j-m}\,a_{j\,-m}$, and
$\tilde{\tilde{a}}_{jm}=-a_{jm}$. The particle number operator
in a given $j$-level is $N_j=\sum_{m}a^{\dagger}_{jm} a_{jm}$. 
 
Three operators,
\begin{equation}
{\cal L}^{-}_j\equiv P_j\,,\quad {\cal L}^{+}_j\equiv P^{\dagger}_j, \quad 
{\cal L}^{z}_j\equiv \frac{1}{2}\left(N_j-\frac{\Omega_j}{2}\right),   \label{1}
\end{equation}
close a quasispin SU(2)
algebra; the corresponding classification of the basis
is very useful in practical calculations.
The quasispin formalism used here reduces the pairing problem to the problem
of coupling of partial quasispins $\vec{{\cal L}_{j}}$
for individual $j$-levels. We use a coupling 
scheme where the basis
states are labeled with the absolute value ${\cal L}_{j}$ of the partial
quasispin, or, alternatively, partial seniority 
$s_j=\Omega_{j}/2-2{\cal L}_{j}$, and particle number $N_{j}$ for
each single-particle level $j$ related to the 
quasispin projection ${\cal L}^{z}_j$. In contrast to the occupancies $N_{j}$,
all partial seniorities $s_{j}$ are conserved by
the pairing Hamiltonian as well as the full seniority of the system
$s=\sum_{j}s_{j}$. The decomposition of the space into blocks labeled by the
seniority quantum numbers simplifies the problem immensely.
Although here we consider only one kind of particles, the 
technique can be extended for full isovector pairing
\cite{hecht65a,hecht65b,volya}.

As a realistic example we consider the shell-model description of
the $^{116}$Sn isotope.
The valence
neutron pairing is well known to play an important role in tin isotopes
\cite{andreozzi96,sandulescu97}. The model space consists of
five single-neutron orbitals
$h_{11/2},\,d_{3/2},\,s_{1/2},\,g_{7/2},$ and $d_{5/2}$ with corresponding
single-particle energies  -9.76, -8.98, -7.33, -7.66, and  -7.57 MeV. There are
601,080,390 many-body states,
among those 272,828 have spin zero and can be partitioned into 420
different seniority sets. The dimension of the fully paired set of seniority
zero (all $s_{j}=0$) is 110. 

In our analysis of the properties of  pure pairing
we set to zero all non-pairing parts of the effective interaction and
use the pairing matrix elements ($L=0$ and $T=1$) from  the 
$G$-matrix derived from the 
nucleon-nucleon interaction \cite{machleidt96}, with
$^{132}$Sn taken as a closed shell \cite{HJ}. In this derivation
the $\hat{Q}$-box method includes all
non-folded diagrams up to the third order in the interaction and sums up the
folded diagrams to infinite order \cite{hjorth95}.
In this way our input
quantities in the Hamiltonian (\ref{BCS:ham}) are completely determined.
Calculations with these parameters, using direct shell-model diagonalization,
as well as the EP+monopole technique \cite{EP} are in a good agreement with
experimental data \cite{YKIS}.
We need to stress however, that our main results and the general discussion are
insensitive to the precise choice of the pairing matrix elements.

\section{Low-lying states and pair vibrations}

The pairing interaction of Eq. (\ref{BCS:ham}) contains the elementary 
processes of pair transfer from one single-particle orbital to another.
Since the pairs are not broken, the interaction
is only capable of mixing the independent particle configurations
within the same seniority set $\{s_{j}\}$. We refer to the amplitude of
such a process as that of a pair
vibration. Pair vibrations were first proposed by 
A. Bohr \cite{bohr68}
addressing the collective nature of low-lying  $0^{+}$ states.
A number of other studies followed \cite{bes66} with a discussion of the 
collective excitations of the pair condensate in the BCS+RPA framework.
The  weakening of the static pairing near the
subshell closure of $^{114}$Sn implies strong fluctuations and
mixing of configurations in this region that may effect the quality of
BCS-based theories in the description of the observed states in Sn nuclei.  

As we mentioned, the total number of the states $J^{\pi}=0^{+}$
in the valence space of $^{116}$Sn
that have no broken pairs (total seniority $s=\sum_{j}s_{j}=0$) equals 110.
The ground state corresponds to the BCS-like coherent
alignment of quasispins in the condensate which produces minimum energy and
average occupancies $N_{j}$ which in the BCS limit would be determined by the
Bogoliubov transformation coefficients, $v_{j}^{2}=\overline{N_{j}},\,\,
u_{j}^{2}=1-\overline{N_{j}} $.
The pair scattering between the shell model basis configurations 
is responsible for their complicated mixing. In what follows we first 
analyze the degree of complexity of the resulting eigenstates with the aid 
of information (Shannon) entropy and invariant correlational (von
Neumann) entropy  \cite{entr,sokolov98,big} and then try to single out the
collective pair vibration modes comparing them with the random phase
approximation (RPA).

\subsubsection{Complexity of the eigenstates}

Information entropy $I_{\alpha}$ of the eigenstates $|\alpha\rangle$ with
respect to the basis $|k\rangle$ was repeatedly used for characterizing the
degree of mixing and delocalization of the eigenfunctions expressed in the
original ``unperturbed" basis \cite{entr,Yon,Reichl,Izr,MF}.
As discussed in \cite{big,ann},
the annoying basis dependence of this quantity may be advantageous providing an
additional knowledge about the dynamic 
interrelation between the reference basis and
the eigenbasis of the Hamiltonian. For example, it was shown recently
\cite{HBZ} that for a shell-model system with {\sl random} 
two-body interactions,  
information entropy of the eigenstates expressed in the basis of the
stationary states generated by the realistic effective interaction does not
evolve along the spectrum being steadily close 
to the random matrix theory limit. This means that the wavefunctions of the
system with random interactions are delocalized all over the spectrum of the
realistic eigenstates. Contrary to that, in the mean-field (shell-model) basis
both sets of the eigenstates display the regular energy dependence typical for
the strong mixing at high level density. 

\begin{figure}
\begin{center}
\epsfxsize=17.0cm \epsfbox{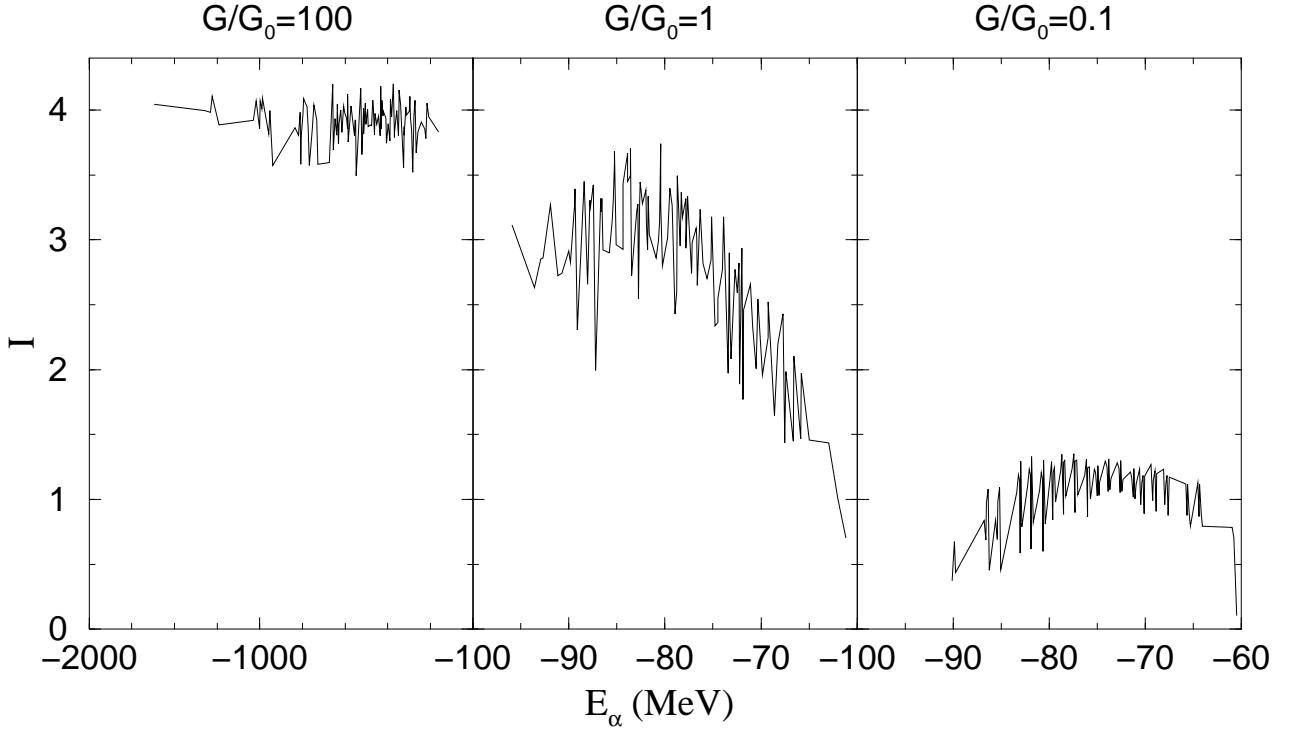}
\end{center}
\caption{Information entropy of seniority zero states in $^{116}$Sn 
plotted versus
their energy. The middle panel corresponds to the realistic values of the
pairing matrix elements, while for the left panel the pairing is
enhanced by a factor of 100, and for the right panel the
pairing is reduced by a factor of 10.
\label{Sn116_info_entr}}
\end{figure}

In Fig. \ref{Sn116_info_entr}, the information entropy
of all 110 states with $s=0$ is shown as a function of state energy
$E_{\alpha}$.
Information entropy of an eigenstate $|\alpha\rangle$,
\begin{equation}
I_{\alpha}=-\sum_k |C^\alpha_k|^2\,\ln (|C^\alpha_k|^2),    \label{2}
\end{equation}
is defined through its decomposition 
in terms of the basis components $|k\rangle$,
\begin{equation}
|\alpha\rangle=\sum_\alpha C^\alpha_k\, |k\rangle.            \label{3}
\end{equation}
Each basis state $|k\rangle$ corresponds to a distribution of particle pairs
over single-particle levels in such a  way that each level has
an even number of particles and they are all paired. The basis wave
functions for 110 different pairwise rearrangements of 16 valence neutrons
in the $^{116}$Sn shell-model space are explicitly built by the action of the 
operators $(P^{\dagger}_{j})^{N_{j}}$ and correctly normalized taking 
into account the blocking effects. 

Three panels in Fig. \ref{Sn116_info_entr} show the entropy $I_{\alpha}$
as a function of energy $E_{\alpha}$ for various pairing strengths measured
with the aid of the overall scaling factor $G/G_{0}$ in pair transfer matrix
elements. The results  are not symmetric with respect to the centroid of the
spectrum since the states of $s=0$ comprise only a part of all states with spin
$J=0$; the remaining states have seniority 4 and higher. 
Average information entropy of states
in the extreme limit of randomness given by the
Gaussian Orthogonal Ensemble (GOE) is predicted to be \cite{Izr,Jones,big}
\begin{equation}
I_{\rm GOE}=\ln (0.482 {\cal N}),                          \label{4}
\end{equation}
where ${\cal N}$ is the space dimension. In our case of
${\cal N}=110$, the GOE limit would give $I_{\rm GOE}=3.97.$ As seen from Fig.
\ref{Sn116_info_entr}, the complexity of states increases with the pairing
strength and reaches the GOE limit for strong pairing. Already 
the realistic pairing strength provides strong mixing of 
all states with some reduction towards the ends of the spectrum.
The asymmetry of $I_\alpha$ as a function of $E_\alpha$ is related to the
asymmetry of the single-particle energies in the shell.
The highest states mainly correspond to the occupation of the highest orbits
(``negative temperature''). Because of the relatively
large energy gap between these
levels and the remaining (lowest) unoccupied ones $h_{11/2}$ and $d_{3/2}\,,$
the pairing interaction is
less effective in state mixing. 

Being a measure of the delocalization of the eigenstates in an unperturbed
basis, information entropy cannot distinguish between a coherent superposition 
of simple excitations characteristic for collective modes and a chaotic
combination which emerged from complicated incoherent dynamics
\cite{big}. (The distinction would require different tools, 
such as invariant correlational entropy \cite{sokolov98}, see the
discussion of Fig. \ref{two_level} below). 
Fig. \ref{Sn116_ent_of_g_100} demonstrates the behavior of information
entropy at large
pairing strengths. The highest excited state approaches the GOE limit very
slowly since, being formed as the ``antipaired" combination at weak pairing, it
is shifted up already in the second order of perturbation theory so that the 
further mixing becomes suppressed. The similar effect of the shift down exists
for the ground state as well but here the collectivity significantly
increases the value of entropy.
The ground state entropy is saturated even above
the GOE limit indicated in the figure by a thin solid line. For very strong
pairing, single-particle energies can be neglected turning the problem into
the degenerate model case, for ${\cal V}_{j\,j'}=\,{\rm const}\,;$ the
realistic interaction is close to this limit. In terms of quasispin, different
recoupling schemes can be used to construct the basis \cite{kerman61}. The basis
states used in EP are labeled both
by the partial quasispins and their projections 
appropriate for the weak coupling
limit with the distribution of particles close to the Fermi
step function. The basis of total quasispin  and different recoupling quantum 
numbers may
serve better in the strong pairing (degenerate) limit. In terms of the particle
distribution, the strong pairing limit with ${\cal V}_{j\,j'}=\,{\rm const}\,$
results in the equipopulation of the sublevels (``infinite
temperature"). 

\begin{figure}
\begin{center}
\epsfxsize=7.0cm \epsfbox{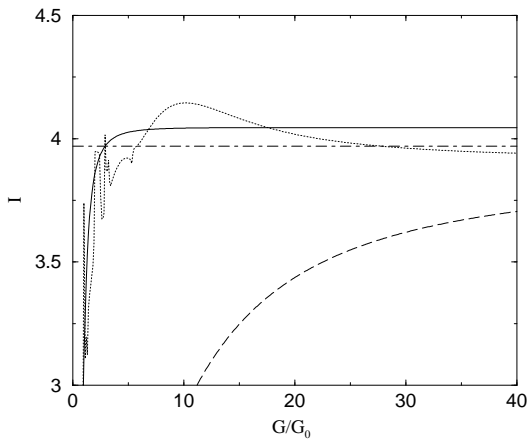}
\end{center}
\caption{Information entropy of selected seniority zero states in $^{116}$Sn
as a function of the pairing strength.
The chosen states are ground state (solid line),
50th state (in order by energy, dotted line) and
the highest, 110th, state (dashed line).
\label{Sn116_ent_of_g_100}}
\end{figure}

At low pairing strengths (the case of $G/G_0=0.1$ in Fig. 
\ref{Sn116_info_entr}), the quality of the occupation
number basis is improved due to the increased role of the mean-field energies.
Here the BCS theory would show no condensate formation since 
the spacings between the
single-particle levels are greater than the pairing strength
by an order of magnitude.
However, in the exact solution, the pair
fluctuations still provide 
a considerable mixing that persists throughout the spectrum. 

\section{Random phase approximation in the quasispin formalism}

\subsubsection{RPA equations}

Our exact solution obtained by a direct diagonalization in the small space of
seniority $s=0$ can be compared with the standard BCS solution
supplemented by the random phase approximation (RPA) for the collective modes
related to pair vibrations. We start with formulating the BCS+RPA approach in
the quasispin formalism. For macroscopic superconductors (with all partial
spins equal to $j=1/2$) such an approach was
used in the classic paper by Anderson \cite{anderson58}. Since the particle number
in this approach is assumed to be preserved only on average, one needs to
introduce the chemical potential $\mu$.

The pairing field can be written as a linear functional of one of the quasispin
operators ${\cal L}$,
\begin{equation}
\Delta_{j}\{{\cal L}\}= \sum_{j'}{\cal V}_{jj'}{\cal L}_{j'}.    \label{5}
\end{equation}
The operator equation of motion for the partial quasispin components with the
pairing Hamiltonian are
\begin{equation}
[{\cal L}_{j}^{x},\,H]=-2i\epsilon_{j}'{\cal L}_{j}^{y}+i[{\cal L}_{j}^{z},
\Delta_{j}\{{\cal L}_{j}^{y}\}]_{+},                        \label{x}
\end{equation}
\begin{equation}
[{\cal L}_{j}^{y},\,H]=2i\epsilon_{j}'{\cal L}_{j}^{x}-i[{\cal L}_{j}^{z},
\Delta_{j}\{{\cal L}_{j}^{x}\}]_{+},                        \label{y}
\end{equation}
\begin{equation}
[{\cal L}_{j}^{z},\,H]=i[{\cal L}_{j}^{y},\Delta_{j}\{{\cal L}_{j}^{x}\}]_{+}
-i[{\cal L}_{j}^{x},\Delta_{j}\{{\cal L}_{j}^{y}\}]_{+},         \label{z}
\end{equation}
where $[A,B]_{+}\equiv AB+BA$ denotes the anticommutator, and we introduced the
corrected single-particle energies $\epsilon_{j}'=\epsilon_{j}+
{\cal V}_{jj}/2-\mu$. 

The BCS solution corresponds to the static mean field alignment of partial 
quasispins; the spread of the single-particle energies plays the role of the
magnetic field along the $z$-axis. The direction of alignment in the $xy$-plane
is chosen arbitrarily (spontaneous symmetry breaking); for instance, we take 
it in such a way that $\langle {\cal L}_{j}^{y}\rangle=0$. 
The static solution can
be derived from the diagonal matrix elements of the equations of motion. We
denote the static amplitudes of ${\cal L}_{j}^{z}$ and ${\cal L}_{j}^{x}$ as
$Z_{j}$ and $X_{j}$, respectively. Linearizing the equations of motion for
off-diagonal matrix elements, we obtain normal modes of pair vibrations in the
harmonic approximation. It is convenient to introduce collective coordinates
$\alpha$ and corresponding conjugate momenta $\pi$ and look for a solution of
the form 
\begin{equation}
{\cal L}_{j}^{x}=X_{j}+x_{j}\alpha+..., \quad {\cal L}_{j}^{y}=y_{j}\pi + ...,
\quad {\cal L}_{j}^{z}=Z_{j}+ z_{j}\alpha+...                    \label{6}
\end{equation}
(in the harmonic approximation the distinction between the coordinates
and momenta is a matter of convention).
Searching for the excitation in a given nucleus, we need to impose the 
average particle number conservation condition. With the definition (\ref{1})
of the operator ${\cal L}_{j}^{z}$, the total particle number operator is
\begin{equation}
N=\sum_{j}\left(\frac{\Omega_{j}}{2}+2{\cal L}_{j}^{z}\right)
\equiv \frac{\Omega}{2}
+2{\cal L}^{z}.                                                  \label{6a}
\end{equation}
In the operator expansion (\ref{6}) this means that the static and oscillating
amplitudes of ${\cal L}_{j}^{z}$ should satisfy 
\begin{equation}
N=\frac{\Omega}{2}+2\sum_{j}Z_{j}, \quad \sum_{j}z_{j}=0.       \label{6b}
\end{equation}

The BCS equation for the 
static part is to be found from the expectation value of the right hand 
side of the $y$-equation (\ref{y}),
\begin{equation}
X_{j}=\frac{\Delta_{j}\{X\}}{\epsilon_{j}'}Z_{j}.              \label{7}
\end{equation}
The normalization of the static amplitudes comes from the condition for the
maximum value of the total quasispin in the ground state,
\begin{equation}
X_{j}^{2}+Z_{j}^{2}=\vec{{\cal L}_{j}}^{2}=\frac{\Omega_{j}}{4}
\left(\frac{\Omega_{j}}{4}+1\right)
\approx \left(\frac{\Omega_{j}}{4}\right)^{2}.                  \label{8}
\end{equation}
The solution can be written now in the form
\begin{equation}
X_{j}=-\frac{\Omega_{j}}{4e_{j}}\Delta_{j}\{X\}, \quad 
Z_{j}=-\frac{\Omega_{j}}{4e_{j}}\epsilon_{j}',                  \label{9}
\end{equation}
where the quasiparticle energies are $e_{j}=+\sqrt{\epsilon_{j}'^{2}+
\Delta_{j}^{2}\{X\}}$, and the gap equation for $\Delta_{j}\equiv
\Delta_{j}\{X\}$ is the same as in the BCS theory,
\begin{equation}
\Delta_{j}=-\sum_{j'}{\cal V}_{jj'}\Omega_{j'}\frac{\Delta_{j'}}{4e_{j'}}.
                                                              \label{10}
\end{equation}
The first condition (\ref{6b}) of particle number conservation determines the
chemical potential in full agreement with the BCS where the occupation factors
are $v_{j}^{2}=(1/2)[1-(\epsilon_{j}'/e_{j})]$.

Using the harmonic oscillator collective Hamiltonian $H=(\pi^{2}+\omega^{2}
\alpha^{2})/2$ 
with frequency $\omega$, we obtain for the RPA amplitudes $x_{j},y_{j}$ and 
$z_{j}$ defined in eq. (\ref{6}):
\begin{equation}
\omega^{2}x_{j}=-2\epsilon_{j}'y_{j}+2Z_{j}\Delta_{j}\{y\},   \label{11}
\end{equation}
\begin{equation}
-y_{j}=2\epsilon_{j}'x_{j}-2Z_{j}\Delta_{j}\{x\}-2\Delta_{j}z_{j},
                                                                 \label{12}
\end{equation}
\begin{equation}
\omega^{2}z_{j}=2\Delta_{j}y_{j}-2X_{j}\Delta_{j}\{y\}.          \label{13}
\end{equation}
The spurious mode of this set, $\omega=0$, gives, either from eq. (\ref{11})
or from eq. (\ref{13}), the solution for $y_{j}$ coinciding with that for 
the static
amplitude $X_{j}$, eq. (\ref{9}). This obviously corresponds to the Goldstone
rotation of the condensate phase. The spurious mode related to the change of
the particle number is generated by the operator $\alpha\sum_{j}z_{j}$, see the
second condition in Eq. (\ref{6b}). However it does not contribute to physical
modes since the sum over $j$ in Eq. (\ref{13}) identically vanishes due to the
symmetry of the kernel ${\cal V}_{jj'}$.
Looking for the non-spurious solutions with
$\omega\neq 0$ and 
eliminating the coordinate amplitudes $x_{j}$ and $z_{j}$, we come to the 
equation for the momentum amplitude $y_{j}$ which determines the physical 
normal modes,
\begin{equation}
(4e_{j}^{2}-\omega^{2})y_{j}=-e_{j}\Omega_{j}\Delta_{j}\{y\}
+2Z_{j}\Delta_{j}\{t\},      \label{14}
\end{equation}
where the last term contains the same functional $\Delta$, Eq. (\ref{5}), 
taken for
\begin{equation}
t_j=\epsilon_{j}'\left(2y_{j}+\frac{\Omega_{j}}{2e_{j}}\Delta_{j}\{y\}\right). 
                                                                \label{15}
\end{equation}
This secular equation is valid also for the normal
modes generated by pairing with no condensate present when we have only
the trivial BCS solution, $X_{j}=0, \Delta_{j}=0$, the
vibrational amplitude $z_{j}$ vanishes, and the single-particle energies are
equal to $|\epsilon_{j}'|$. With the extension to higher orders of the same 
mapping procedure to the collective variables as in Eq. (\ref{6}), one can
study the anharmonic effects. However, if the anharmonicity is indeed
important, as for example in the region of very low frequencies and
correspondingly large amplitude collective motion, it is simpler (at least in
the pure pairing problem) to switch to the exact solution.

\subsubsection{Example: two-level system with off-diagonal pairing}

As an example of the analytically solved RPA, 
we consider an interesting particular case 
of two levels with capacities $\Omega_{1,2}$ and single-particle energies 
$\epsilon_{1,2}$ when the pairing interaction has only the off-diagonal
amplitude ${\cal V}_{12}={\cal V}_{21}\equiv g$ (in this case the sign of $g$
does not matter). With the effective interaction parameters $\lambda_{1,2}=
g\Omega_{1,2}/4$, the set of the BCS gap equations (\ref{10}) takes the form
\begin{equation}
\Delta_{1}=\frac{\lambda_{2}}{e_{2}}\Delta_{2}, \quad
\Delta_{2}=\frac{\lambda_{1}}{e_{1}}\Delta_{1},                \label{16}
\end{equation}
which leads to the exact solution
\begin{equation}
\Delta_{1}^{2}=\frac{\lambda_{1}^{2}\lambda_{2}^{2}-{\epsilon^{\prime}_{1}}^{2}
{\epsilon^{\prime}_{2}}^{2}}{\lambda_{1}^{2}+{\epsilon^{\prime}_{2}}^{2}}, \quad
\Delta_{2}^{2}=\frac{\lambda_{1}^{2}\lambda_{2}^{2}-{\epsilon^{\prime}_{1}}^{2}
{\epsilon^{\prime}_{2}}^{2}}{\lambda_{2}^{2}+{\epsilon^{\prime}_{1}}^{2}}.         \label{17}
\end{equation}
The corresponding quasiparticle energies are given by
\begin{equation}
e_{1}^{2}=\lambda_{1}^{2}\frac{\lambda_{2}^{2}+{\epsilon^{\prime}_{1}}^{2}}
{\lambda_{1}^{2}+{\epsilon^{\prime}_{2}}^{2}}, \quad
e_{2}^{2}=\lambda_{2}^{2}\frac{\lambda_{1}^{2}+{\epsilon^{\prime}_{2}}^{2}}
{\lambda_{2}^{2}+{\epsilon^{\prime}_{1}}^{2}},                             \label{18}
\end{equation}
with the useful identity 
\begin{equation}
e_{1}e_{2}=\lambda_{1}\lambda_{2}                          \label{18a}
\end{equation}
being valid.
The BCS solution collapses, $\Delta_{1,2}\rightarrow 0$, at the critical 
coupling strength determined by $\lambda_{1}^{2}\lambda_{2}^{2}=
{\epsilon^{\prime}_{1}}^{2}{\epsilon^{\prime}_{2}}^{2}$, or at
\begin{equation}
g^{2}\rightarrow g_{c}^{2}=\frac{16|{\epsilon^{\prime}_{1}}{\epsilon^{\prime}_{2}}|}
{\Omega_{1}\Omega_{2}}.                                         \label{19}
\end{equation}
The secular equation, see (\ref{14}) and (\ref{16}), with the help of the 
identity (\ref{18a}), gives, along with the spurious mode $\omega^{2}=0$, 
the physical root
\begin{equation}
\omega^{2}=4(e_{1}^{2}+e_{2}^{2}+2\epsilon_{1}'\epsilon_{2}')=
4[(\epsilon_{1}'+\epsilon_{2}')^{2}+\Delta_{1}^{2}+\Delta_{2}^{2}]. \label{20}
\end{equation}

\subsubsection{Exact solution versus BCS+RPA}

\begin{figure}
\begin{center}
\epsfxsize=7.0cm \epsfbox{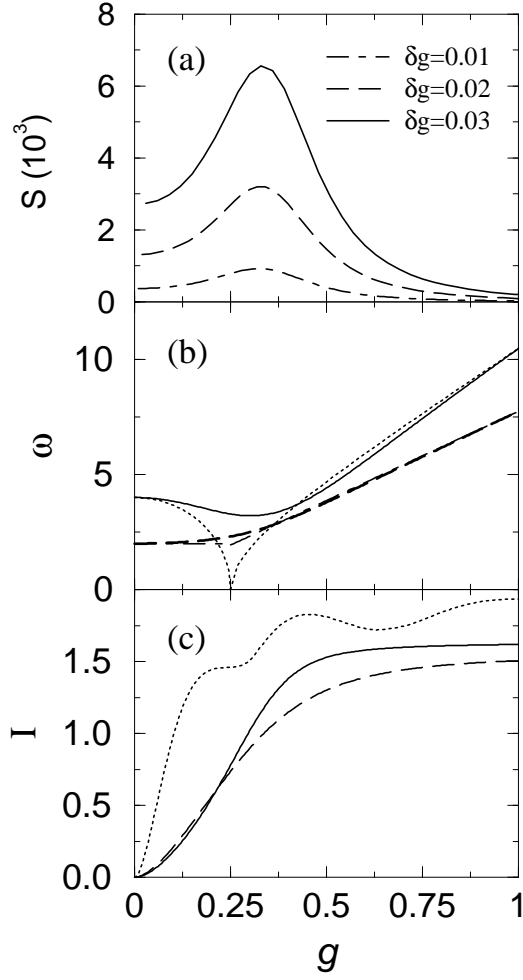}
\end{center}
\caption{Two-level pairing model; $N=\Omega_1=\Omega_2=16\,,$
only the off-diagonal pair transfer
amplitude ${\cal V}_{12}={\cal V}_{21}=g$ is taken into account.
The upper plot (a) displays the invariant entropy, see text,
of the ground state for
$\delta g=0.01,\,\,0.02\,,$ and $0.03$ averaging intervals. 
The middle part  (b) shows the excitation energy
of the lowest pair vibration state, thick
solid line. The thin dotted line approximates this curve with the aid of the
RPA built on
the normal Fermi ground state on the left of the phase transition point at 
$g=0.25$
and on the BCS ground state on the right of the critical point. 
Thick dashed line shows the excitation energy of the lowest state with
broken pair $s=2$. This curve is compared with $2e$, thin dash-dot curve,
the BCS energy of a two-quasiparticle state.
The lower panel (c) displays information entropy for the ground state
(solid line), second pair vibration excited state $s=0$ (dotted line),
and the lowest $s=2$ state (dashed line).   
\label{two_level}}
\end{figure}

\begin{figure}
\begin{center}
\epsfxsize=7.0cm \epsfbox{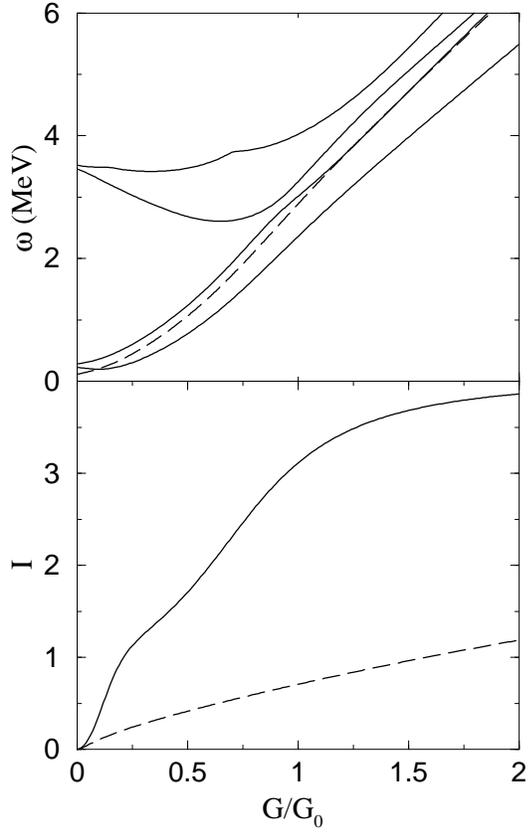}
\end{center}
\caption{Upper panel: 
excitation energies of low-lying states in the shell model for $^{116}$Sn 
as a function of relative pairing strength $G/G_0$,
the lowest pair vibration states with $s=0$, solid
line, and the lowest $s=2$ state, dashed line.
Lower panel: information entropy of the ground state (solid line) and 
and the most excited paired state $s=0$ (dashed line).\\
\label{Sn116_ent_w_g_10}}
\end{figure}

The behavior
of energies and entropy in the BCS phase transition region are illustrated in
Fig. \ref{two_level} for a two-level model and Fig.
\ref{Sn116_ent_w_g_10} for the realistic case.
First we consider the two-level model
of the previous section see Fig. \ref{two_level}.
In the case of  half occupancy and two levels of equal capacity,
$N=\Omega_{1}=\Omega_{2}$, with 
energies $\pm \epsilon$ symmetric with respect to the chemical potential
$\mu=0\,,$
the RPA predicts $\omega^{2}=8\Delta^{2}\,,$ see Eq. (\ref{20}).
Note that in the case of many interacting
levels with ${\cal V}_{jj'}=\,$const the spectrum of normal modes starts with 
the lower value $\omega^{2}=4\Delta^{2}$ (threshold of pair breaking).

The panel (b) in the middle shows the energies of the
lowest pair vibration state
(thick solid line) and the lowest state with one broken pair (thick
dashed line) as a function of the pairing strength $g$.
The excitation energy
of the pair vibration is compared with the RPA prediction shown by the
thin dotted line. The BCS phase transition occurs, in the units corresponding 
to $|\epsilon_{1,2}|=1$, at $g=g_{c}=0.25$, where we see  
the breakdown of the RPA and the vanishing collective frequency.
Below this point the RPA is constructed on the background of the normal Fermi 
distribution, while above $g=g_{c}$ the
RPA is built on the pairing condensate.  
The limits $g\rightarrow 0$ and $g\rightarrow\infty$ are described
well within the RPA while near the phase transition large fluctuations make 
the BCS+RPA description unreliable. The failure of the BCS
is well discussed in the literature and studied using simple models
\cite{bang70}. In accordance with this instability,
a sharp rise of entropy takes place
in the vicinity of $g=g_{c}$ showing a dramatic
restructuring of the eigenstates when the independent pair basis states
become strongly mixed. One needs to emphasize here that, even for a 
very strong interaction, the limit of the degenerate seniority model 
is never reached if the pair transfer matrix elements are different for
different pairs, in contrast to the model with constant pairing when all
${\cal V}_{jj'}=g=$constant.

Fig. \ref{two_level}(b) addresses the
question of the nature of the first excited state. The breaking of
a pair  requires two-quasiparticle excitation energy of $2e$
that is plotted with a thin dashed line,
which has in the asymptotic limit of strong 
pairing the value of twice the BCS gap. 
In realistic nuclear systems unpaired particles
may still interact attractively therefore reducing this energy even further.
The lowering of a two-quasiparticle quadrupole mode
is especially strong
in the case of collective vibrations near the onset of deformation.
The lowest state of pair vibrations, which has  zero spin, is usually much
higher in energy.
It is
possible in principle that the state of pair vibrations becomes the lowest
excited state. For example this can be the case when two single-particle
orbitals lie very close in energy making a pair transfer energetically
more favorable then breaking  a pair.
This feature is not present in the two-level model
and the lowest thick dashed line, corresponding to the $s=2$ state
in the plot in Fig. \ref{two_level}(b),
is always lower than the thick solid line representing the pair vibrations.

\begin{figure}
\begin{center}
\epsfxsize=7.0cm \epsfbox{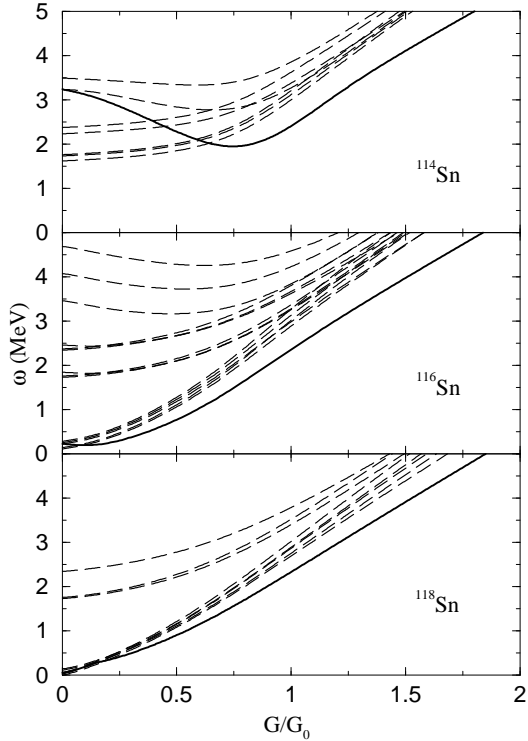}
\end{center}
\caption{Excitation energies of low-lying $s=2$ states (dashed lines) are
compared with that of
the lowest pair-vibration state $s=0$ shown (solid line) as a function of the
relative pairing strength. The three panels
correspond to $^{114}$Sn, $^{116}$Sn, and $^{118}$Sn isotopes.
\label{Sn_w_g}}
\end{figure}

\begin{figure}
\begin{center}
\epsfxsize=7.0cm \epsfbox{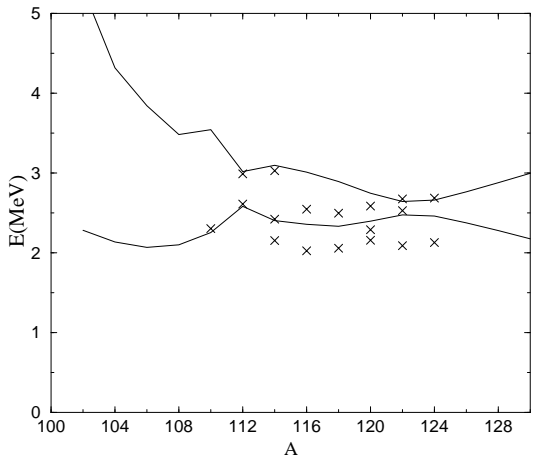}
\end{center}
\caption{Excitation energies of the lowest two pair vibration states
are plotted as a function of the nuclear mass $A$ for Sn isotopes.
Experimentally measured energies of the lowest spin-$0$ states
that are expected to have strong $s=0$ components,  are shown
for comparison.
\label{Sn_0P_PV}}
\end{figure}

In the $^{116}$Sn model, the
upper plot of Fig. \ref{Sn116_ent_w_g_10}, the lowest
excited $0^{+}$ state is of a pair-vibration nature. This pattern
persists through the entire range of couplings displayed. The same trend is
observed in other Sn isotopes, Fig. \ref{Sn_w_g}, including the case of
$^{114}$Sn with $G/G_0$ above 0.7. Here subshell closure requires higher energy
for the $s=0$ pair vibration which in the limit of weak pairing is just a
two-particle-two-hole excitation of the normal Fermi ground state. It comes
from the fact that there exist close single-particle
levels that are not fully occupied
and thus  opened for low energy pair vibrations. In $^{116}$Sn these are 
the $h_{11/2},\,d_{3/2},$ and $s_{1/2}$ levels, which are separated by only
a few hundred keV. Because of the presence of other
interactions, this prediction for the first excited state is
not fulfilled in reality. Still, a number of
low-lying $0^{+}$ states is observed in Sn isotopes. 
The nature of these states is
complex being influenced by other phenomena such as
proton core excitation and shape coexistence \cite{woods92}.
Moreover, the
effect of other interactions can be important through the admixture of
seniority $s=4$ states that may have collective structure of a two-phonon
excitation. 
In Fig. \ref{Sn_0P_PV} excitation energies of the
two lowest pair vibration states
are compared with the experimentally observed states that are expected to
be mostly of a pair vibration nature, see \cite{woods92} and references therein.
The agreement is satisfactory considering the crudeness of the pure pairing
model and the uncertainties in the pair transfer strengths and
single-particle energies.  

It has already been discussed in the literature that even though pairing is
generally quite strong in Sn isotopes there is some weakening in the region
of $^{114}$Sn due to the nearly 2 MeV shell gap between the 
$h_{11/2},\,d_{3/2}$ and $s_{1/2}$ levels and the rest of the single-particle
orbitals \cite{EP}. The proximity of the BCS phase transition results in a
rapid growth of information entropy in this region, see
Fig. \ref{Sn116_ent_w_g_10}, lower panel. This also leads to a large
excitation energy of the lowest pair-vibration
state in $^{114}$Sn at  relatively
weak pairing, Fig. \ref{Sn_w_g}. The large configuration mixing in the
region of a subshell closure is supported by experimental observations in the
two-neutron pick-up reaction \cite{fleming70} and has a direct manifestation
in the increase of information entropy.

The randomizing aspect of pair vibrations is somewhat limited to the
region near the BCS phase-transition, where many recoupling schemes of
quasispin become equiprobable. A similar phenomenon for angular
momentum is known
as geometric chaoticity \cite{mulhall} where the many-particle states of good
total spin constructed by various recoupling schemes have an equivalent mixing 
(entropy). With no preference
to any particular way of coupling, a minor change in the interaction
may lead to a complete restructuring of the spectrum, a
phenomenon similar to chaoticity. To emphasize the appearance of chaoticity
in the BCS phase-transition region the invariant entropy for a two level model
is presented in Fig. \ref{two_level}(a).
The invariant
entropy (basis independent)
shows the degree of restructuring that happens in the eigenstate in
response to a change $\delta g$ in the interaction parameter $g\,.$
The invariant (von Neumann) entropy of an eigenstate $|\alpha\rangle$ is
determined with the aid of the density matrix $\rho^\alpha$
\begin{equation}
S_\alpha=-{\rm Tr} \left \{\rho^\alpha  \ln \rho^\alpha\right \}\,,
\end{equation}
which is defined as an average over a region of parameters
\begin{equation}
\rho^\alpha_{k\,k'} (g,\delta g)=\frac{1}{\delta g} \int_g^{g+\delta g} \,
{C^{\alpha}_{k'}}^{*} C^{\alpha}_{k}\,.
\end{equation}
The  invariant entropy in Fig. \ref{two_level}(a) is peaked in
the phase transition region, indicating a very sensitive (chaotic)
dependence of an eigenstate on the interaction strength. Similar sharp
enhancements of invariant entropy near the phase transitions has been found in
the interacting boson models \cite{cejnar01}.
\section{Local spectral statistics}
\begin{figure}
\begin{center}
\epsfxsize=7.0cm \epsfbox{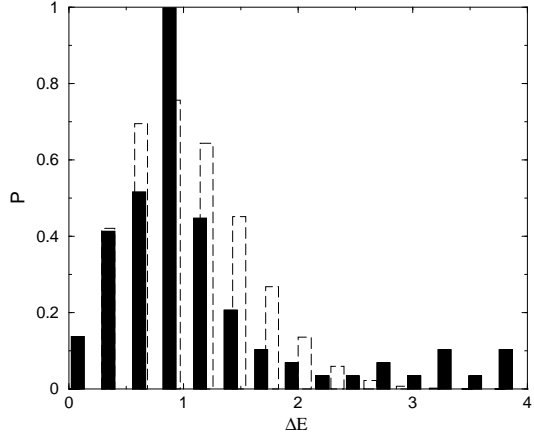}
\end{center}
\caption{The nearest neighbor spacing distribution of $s=0$ states in
$^{116}$Sn shown, filled histogram; the GOE distribution, open columns.
\label{Sn116_PS}}
\end{figure}

\begin{figure}
\begin{center}
\epsfxsize=7.0cm \epsfbox{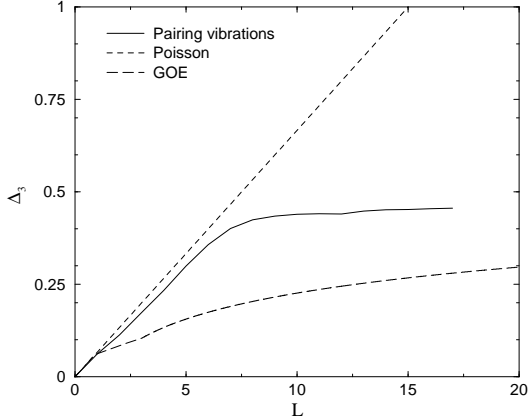}
\end{center}
\caption{The spectral rigidity ($\Delta_3$ statistics) of paired states in
$^{116}$Sn compared to that of the GOE and Poisson distributions.
\label{d3f}}
\end{figure}

The nearest neighbor spacing distribution $P$ shown in Fig. \ref{Sn116_PS} is
very close to that of the GOE exhibiting the features of level repulsion.
However, the tail of the distribution is not Gaussian; it reveals an enhanced
fraction of large gaps between the levels.
The so-called
spectral rigidity, or $\Delta_3$ statistics, is the
more sensitive probe of spectral chaoticity.
Fig. \ref{d3f} indicates a pseudo-random
character of the spectra of the exactly solved pairing problem.
The spectral rigidity is given as a function of $L$, the length of the
fitted interval, whereas  averaging over all starting points of such
(overlapping) 
intervals is assumed. 
In Fig. \ref{d3f} the spectral rigidity
of paired states in $^{116}$Sn is compared with the chaotic (GOE)
distribution and
regular Poisson-type behavior. 
The linear low $L$ behavior of the spectral rigidity is  Poisson-like, 
while for larger intervals, starting from around $L=7$,
the regularities of the spectrum flatten the curve as
an evidence of the oscillatory behavior associated with a regular
quasi-harmonic component in the spectrum. In fact, we are dealing here with the
non-standard situation of a local level repulsion supplemented by the 
presence of a vibrational order.

\section{Properties of pairing through the entire spectrum}

\subsubsection{Level density}

\begin{figure}
\begin{center}
\epsfxsize=7.0cm \epsfbox{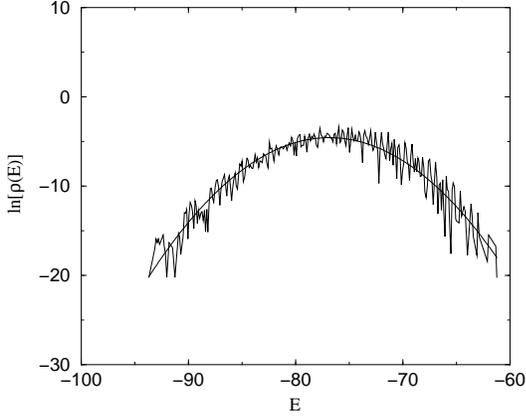}
\end{center}
\caption{Density of states in $^{116}$Sn from the exact solution
with the pairing interaction as a function of energy.
\label{Sn116_entropy}}
\end{figure}

Even if the mixing is strong the amount of states with a
given seniority is very limited. For the lowest
seniority this number is small compared to the total number of states in
the truncated Hilbert space. Globally the level repulsion, 
Fig. \ref{Sn116_PS}, is responsible for the widening of the
entire spectrum. The density of states in the Hamiltonian systems governed by
two-body interactions and described by finite matrices is generically close to
Gaussian (the same is true for the limiting
case of no residual interaction \cite{brody}). Pairing is very typical in this
sense.  Fig. \ref{Sn116_entropy} shows the density of states in $^{116}$Sn
found from the exact solution of the pairing problem. 
The resulting curve is well fit by a Gaussian.

The parameters of the Gaussian are the lowest moments of the Hamiltonian 
which consists of single-particle and pairing parts,
\begin{equation}
H=H_{\rm sp}+\frac{G}{G_0}\,H_{\rm p}\,.
\label{sml_H}
\end{equation}
The centroid of the density distribution is
given by the trace of the Hamiltonian,
\begin{equation}
\overline{H}=\frac{1}{{\cal N}}\, {\rm Tr} (H)\,,          \label{d1}
\end{equation}
where ${\cal N}$ is the total number of states (or levels for the case of
$J=0$) in the shell-model space under consideration.
The width of the level density is the mean square
deviation of the Hamiltonian,
\begin{equation}
\sigma^2=\overline{H^2}-{\overline{H}}^2\,.              \label{d2}
\end{equation}
In general the dependence of the width on the coupling strength $G/G_0$
can contain a linear term
as well as 
a quadratic term. As shown in the
analysis of the statistical properties of the shell-model solutions 
\cite{big}, the two widths, a combinatorial one for noninteracting particles
and a dynamical one due to the residual interaction, are added in quadratures.
This is possible if the two parts of the Hamiltonian (\ref{sml_H}) are
statistically independent, $\overline{H_{\rm sp}}\,\,\overline{H_{\rm p}}= 
\overline{H_{\rm sp}\,H_{\rm p}}$.


\begin{figure}
\begin{center}
\epsfxsize=7.0cm \epsfbox{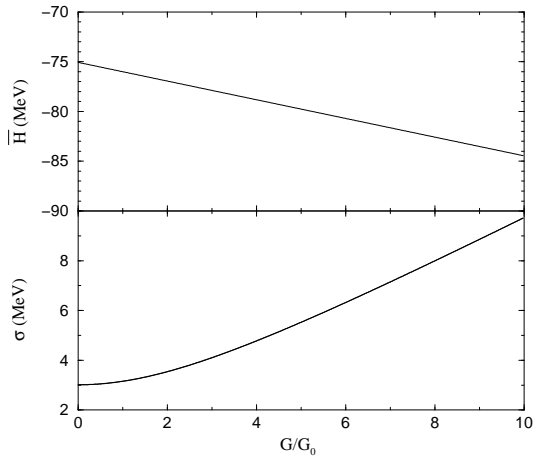}
\end{center}
\caption{The centroid, upper panel, and the width of the density of states,
lower panel, in the pairing model of $^{116}$Sn as a function of the relative 
pairing strength. The empirical curve for the width is
indistinguishable from the fit of Eq. (\ref{d3}).
\label{width_Sn116}}
\end{figure}

In  Fig. \ref{width_Sn116} the mean value and the width of the level density
distribution are shown as functions of the relative pairing strength $G/G_0\,.$
The location of the centroid of the level density  is linear in the coupling
strength,
while the width is exactly fit by
\begin{equation}
\sigma=\sqrt{\sigma_{\rm sp}^2+\left (\frac{G}{G_0} \sigma_{\rm int}\right )^2} 
                                                      \label{d3}
\end{equation}
where the single-particle (non-interaction) width,
$\sigma_{\rm sp}=3.01$ MeV, and the pairing width,
$\sigma_{\rm int}=0.857$ MeV, add  in quadratures. In the region
of the realistic strength the single-particle contribution to the width is
by far the more dominant.

\begin{figure}
\begin{center}
\epsfxsize=7.0cm \epsfbox{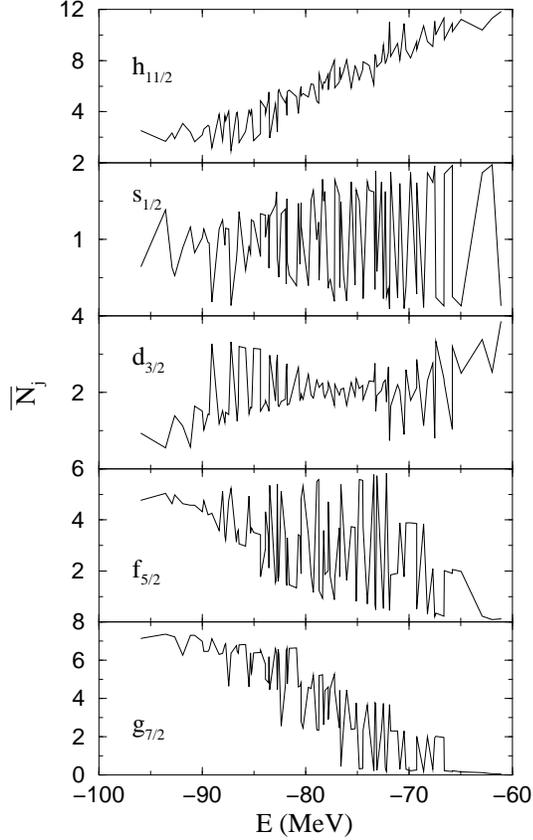}
\end{center}
\caption{Average occupancies of five  single-particle orbitals
for all 110 $s=0$ states in $^{116}$Sn  as a function of 
energy.
\label{s0_statesSn116}}
\end{figure}

\subsubsection{Pairing and thermalization}

A closed mesoscopic system can in principle be analyzed solely in terms of 
individual eigenstates and energy eigenvalues. However, in the region of high
level density, a statistical description operating with thermodynamical
concepts is very appealing. It was shown both for complex atoms 
and nuclei \cite{temp,big} that the increase of chaotic mixing with excitation
energy and level density is in a sense equivalent to
the process of thermalization and can be treated with the help of a
temperature. 
Moreover, the presence of such mixing is necessary in order to make 
the neighboring eigenstates ``look the same" \cite{perc} and justify the
statistical averaging as in a microcanonical ensemble \cite{big,ann}.
In particular, it turned out that the average
occupation numbers of single-particle orbitals can be described by the
Fermi-Dirac distribution even in the presence of strong interactions, although
the single-particle energies can be renormalized to become those of
quasiparticles. Thus, in the statistical region of spectrum we come effectively
to the picture of a heated Fermi-liquid.

However, it is not clear that a specific interaction such as pairing can 
lead to full thermalization.
As shown in Fig. \ref{s0_statesSn116}, the occupancies of single-particle levels
derived for the eigenstates of the pairing problem
vary on average monotonously with energy, therefore showing a trend to
thermalization. At the same time,
there are wild fluctuations which would be even further enhanced
if states of
different seniorities were included. 

To address the thermalization we can
define various temperature scales \cite{temp,big}. The thermodynamical 
``absolute temperature'' comes from the density of states
\begin{equation}
\frac{dS_{\rho}(E)}{dE}=\frac{1}{T_{\rm abs}}\,,\quad {\rm where}\quad
\rho(E)=e^{S_{\rho}(E)}\,;                       \label{d4}
\end{equation}
therefore
\begin{equation}
T_{\rm abs}(E)=\frac{\sigma^2}{\overline{H}-E}\,.
\end{equation}
The ``single-particle" temperature $T_{\rm s-p}$ 
is linked to the termalization of the 
system in terms of the distribution of particles over single-particle levels.
It can be determined for each individual eigenstate from the best fit of the
occupation numbers to the Fermi-Dirac distribution
\begin{equation}
\overline{N}_j=\frac{\Omega_j}{e^{(\epsilon_j-\mu)/T_{\rm s-p}}+1}\,,
\end{equation}                                              \label{d5}
where the chemical potential $\mu$ also can be determined from the two-parameter
fit. 

\begin{figure}
\begin{center}
\epsfxsize=7.0cm \epsfbox{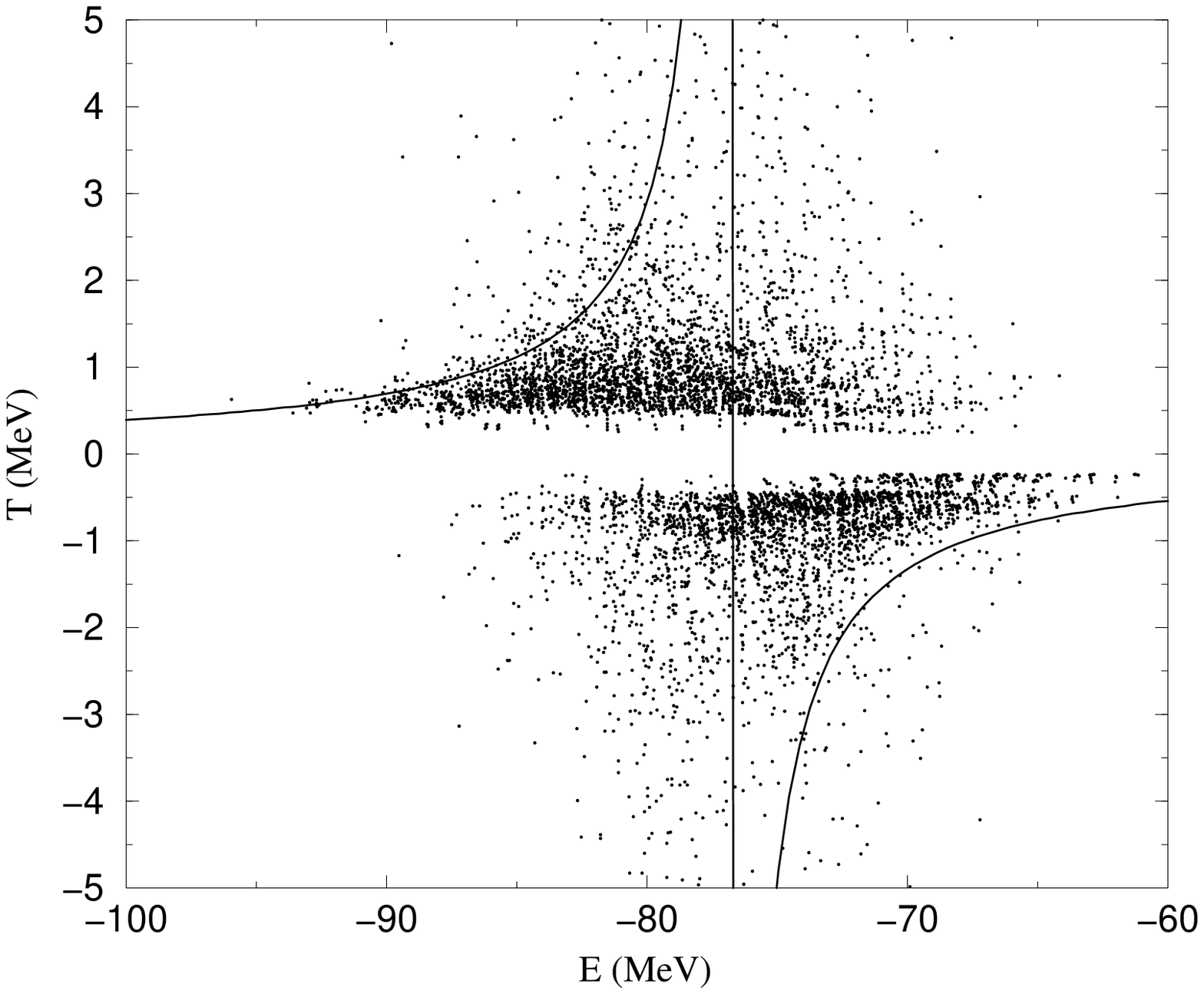}
\end{center}
\caption{Absence of spectral thermalization in
$^{116}$Sn from the comparison of absolute temperature, Eq. (35), solid 
line, with single-particle
temperature that follows from the neutron distribution over single-particle
orbitals, Eq. (37).
\label{Sn116_temprature}}
\end{figure}

In Fig. \ref{Sn116_temprature} the comparison of these two definitions of
temperature for
all states in $^{116}$Sn is shown. The result is very different from what was
observed in the shell model taking into account all residual interactions where
$T_{{\rm abs}}$ and $T_{\rm s-p}$ were in a good agreement for the majority of 
eigenstates (a similar thermalization occurs as well when random residual
interactions are taken). With the pairing interaction only, the
dots representing $T_{\rm s-p}$ of individual
eigenstates show a low temperature along the entire spectrum whereas
the smooth hyperbola corresponds to absolute
temperature which has a singularity in the middle of the spectrum. It is clear
that the single-particle thermometer cannot measure correctly the temperature
of the paired system.


\section{Rotation of a paired system}

A pure pairing fermionic condensate has properties similar to
those observed in macroscopic superfluid systems. Heating, i.e. the
increase of excitation energy, results in a gradual destruction of the
condensate and an increase in the normal component of the fluid (increase 
of seniority in nuclear physics terminology).
Similarly, a higher angular momentum can
be created only by a pair breaking. The unpaired particles can be coupled to 
a nonzero angular momentum which can be treated as some kind of
rotation. When the spins allowed for a given seniority are fully aligned, 
any further increase of angular momentum requires a change of configuration 
and seniority jump, with the corresponding energy increase.
This picture is qualitatively similar to the phenomenon of quantized vortex 
formation in superfluid liquid $^{4}$He. 

\begin{figure}
\begin{center}
\epsfxsize=7.0cm \epsfbox{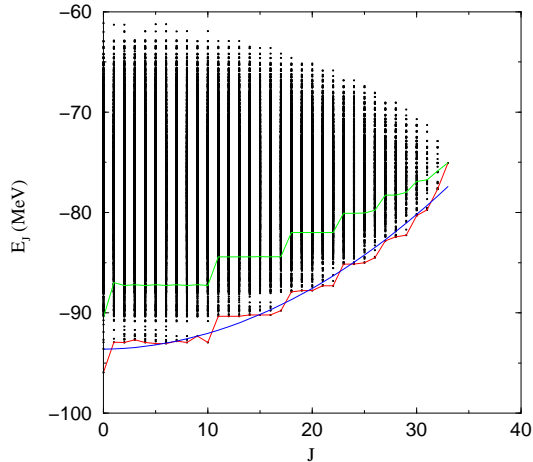}
\end{center}
\caption{Energies of all many-body states in $^{116}$Sn versus
their angular momentum; the yrast line (solid line) is fit by
a parabola (dashed line); the dashed-dotted curve corresponds to the yrast
line in the degenerate case with all single-particle energies being equal.
\label{Sn116_momentum1}}
\end{figure}

In Fig. \ref{Sn116_momentum1} the 
energies of all eigenstates in $^{116}$Sn within the shell model space
(all possible seniorities)
are marked by points in the energy versus angular momentum plane.
The stair-case yrast line, shown by a solid line in Fig \ref{Sn116_momentum1},
is drawn through the states of maximum spin for each seniority. It
can be fitted well by a parabola (dashed line),
thus giving the average moment of inertia  $I=32$ MeV$^{-1}\,.$ This result
can be compared with the classical moment of inertia of a rigid 
sphere with uniform density, $I_{\rm rig}=2 M R^2/5$ with mass $M=938 A$ MeV
and radius
$R=1.2 A^{1/3}$ fm; for $^{116}$Sn  $I_{\rm rig}=35$ MeV$^{-1}\,.$
It is known
that in the adiabatic (linear response) approximation a cranked 
Fermi gas has a
rigid body moment of inertia while pairing interactions usually reduce
this value \cite{Bel,BM} by about a factor of 2. The above results show that
the overall effect of pairing on the rotational properties is considerably
quenched at  high excitation energy. The broken pairs and pair vibrations
return  the moment of inertia to the original Fermi-gas value. This phenomenon 
has a direct analog in superfluid liquid $^{4}$He, where
a large number of vortices emerges at high rotation speed forming a lattice
structure \cite{lane61} which rotates as a whole restoring the rigid body
moment of inertia and characteristic average velocity field.

Pairing vibrations are primarily responsible
for mixing of states in paired systems and therefore they are expected to
have a profound effect on the properties of nuclear systems such as
the high-spin moment of
inertia. For calculating the yrast line shown on Fig. \ref{Sn116_momentum1}
as a dashed-dotted line, we assume full degeneracy, making all $\epsilon_j$ 
equal. This reduces energy required for the pair transfer and therefore makes 
pair vibrations stronger and more chaotic.
As a result, the moment of inertia was even further
increased to a value of 43 MeV$^{-1}\,.$ As could be expected in the 
degenerate model, where the perturbative cranking approach is not valid, the
average angular momentum dependence on energy is vibrational (close to linear),
and the
quadratic contribution related to the inverse moment of inertia is small.

\section{Conclusion}

In this paper we discussed the properties of the pairing interaction in a
finite quantum system described within the shell-model approach. This
interaction, being a very important part of the general residual interaction
 has specific features being effective for pairs with total spin
zero only. The standard approach takes into account the pairing effects with the
aid of the BCS approximation borrowed from theory of macroscopic
superconductivity. This approximation fails not only from the viewpoint of
particle number nonconservation which might be essential for finite systems,
but it also
does not work for weak pairing and beyond the phase transition point.
Therefore we base our study on the earlier suggested exact numerical solution 
where the use of the quasispin symmetry provides necessary quantum numbers and
simplifies the problem immensely. 

The collective excitations in the paired system (pair vibrations) are
traditionally described by developing the random phase approximation on the
background of the BCS ground state. We show that the exact solution differs
from that in the BCS+RPA theory, especially in the region of the phase
transition. It is seen clearly in the example of the realistic shell model for 
the $^{116}$Sn nucleus as well as in the analytically solvable two-level model.
Analogous results can be obtained for isospin-invariant pairing which will be
considered elsewhere.

For the first time we have discussed the chaotic (incoherent) aspects of the
pairing interaction. The residual interactions in general mix independent
particle configurations creating complicated stationary many-body states.
These effects influence both local and global statistical properties of the
system. In the case of a generic residual interaction, we come locally to the 
GOE-like nearest level spacing distribution and enhanced spectral rigidity
whereas the global behavior reveals thermalization of the system. Typically,
the interaction plays the role of a heat bath bringing the single-particle
occupancies close to the Fermi-Dirac distribution even in a strongly
interacting system. The pairing interaction is in many respects exceptional.
Although it induces the local level repulsion and increases information entropy
of the eigenstates, the spectral rigidity shows
a pseudo-oscillatory behavior related to the vibrational character of
excitations. Both information (Shannon) and invariant (von Neumann) entropies
distinctly reflect the existence of the phase
transition. The normal behavior of thermodynamic entropy associated with the
level density coexists with the absence of full thermalization of
single-particle motion. The yrast line of the system reveals the trend to the
rigid-body rotation at high excitation energy where many pairs are effectively
broken. 

Pure pairing cannot exist as the only two-body interaction (for a charged
system it would violate the gauge invariance, see also \cite{cohfluc}).
As shown by the full analysis of the statistical properties of the nuclear
shell model wave functions \cite{big}, 
the pairing can be considerably modified by the
presence of other parts of the residual interaction. For example, the families
of states corresponding to various values of the seniority quantum number are
almost completely mixed although the pairing phase transition is still observed
through the behavior of the pairing correlators for individual low-lying
eigenstates \cite{arg}. However, beyond the phase transition there exists 
a long exponentially decreasing tail of dynamic pairing correlations. The
interplay of pairing with other types of residual interaction is an
interesting and promising problem for future research.

We thank M. Hjorth-Jensen for providing the renormalized $G$-matrix for the
$^{132}$Sn region.
The work was supported by the NSF grant 0070911.

\end{document}